\documentclass[sigconf, authorversion, nonacm]{acmart}

\copyrightyear{2025} 
\acmYear{2025} 
\setcopyright{rightsretained}
\acmConference[GLSVLSI '25]{Great Lakes Symposium on VLSI 2025}{June
30-July 2, 2025}{New Orleans, LA, USA}
\acmBooktitle{Great Lakes Symposium on VLSI 2025 (GLSVLSI '25), June 30-July
2, 2025, New Orleans, LA, USA}
\acmDOI{10.1145/3716368.3735158}
\acmISBN{979-8-4007-1496-2/2025/06}

\usepackage{amsmath,amsfonts}
\usepackage{algorithmic}
\usepackage{graphicx}
\usepackage{textcomp}
\usepackage{xcolor}
\usepackage{cleveref}
\usepackage{subfig}
\usepackage{multirow}

\Crefformat{figure}{#2Fig.~#1#3}

\usepackage[colorinlistoftodos,prependcaption]{todonotes}

\def\BibTeX{{\rm B\kern-.05em{\sc i\kern-.025em b}\kern-.08em
    T\kern-.1667em\lower.7ex\hbox{E}\kern-.125emX}}

\begin{document}

\title{Systolic Arrays and Structured Pruning Co-design \\for Efficient Transformers in Edge Systems}

\author{Pedro Palacios}
\email{pedro.palaciosalmendros@epfl.ch}
\affiliation{%
  \department{Embedded Systems Laboratory (ESL)}
  \institution{École Polytechnique Fédérale de Lausanne (EPFL)}
  \city{Lausanne}
  \country{Switzerland}
}

\author{Rafael Medina}
\email{rafael.medinamorillas@epfl.ch}
\affiliation{%
  \department{Embedded Systems Laboratory (ESL)}
  \institution{École Polytechnique Fédérale de Lausanne (EPFL)}  \city{Lausanne}
  \country{Switzerland}
}

\author{Jean-Luc Rouas}
\email{jean-luc.rouas@labri.fr}
\affiliation{%
  \department{LaBRI, CNRS UMR 5800}
  \institution{Univ. Bordeaux, Bordeaux INP}
  \city{Talence}
  \country{France}
}

\author{Giovanni Ansaloni}
\email{giovanni.ansaloni@epfl.ch}
\affiliation{%
  \department{Embedded Systems Laboratory (ESL)}
  \institution{École Polytechnique Fédérale de Lausanne (EPFL)}
  \city{Lausanne}
  \country{Switzerland}
}

\author{David Atienza}
\email{david.atienza@epfl.ch}
\affiliation{%
  \department{Embedded Systems Laboratory (ESL)}
  \institution{École Polytechnique Fédérale de Lausanne (EPFL)}  \city{Lausanne}
  \country{Switzerland}
}

\renewcommand{\shortauthors}{Palacios et al.}

\begin{abstract}
Efficient deployment of resource-intensive transformers on edge devices necessitates cross-stack optimization.
We thus study the interrelation between structured pruning and systolic acceleration,
matching the size of pruned blocks with the systolic array dimensions. In this setting, computations of pruned weight blocks can be skipped, reducing run-time and energy consumption, but potentially impacting quality of service (QoS).
To evaluate the trade-offs between systolic array size and sparsity opportunities, we present a novel co-design framework that integrates algorithmic optimization, system simulation, and hardware design.
Targeting speech recognition and machine translation using transformers as case study, we analyze how configuration choices across the stack affect performance metrics.
Results demonstrate that structured pruning on systems featuring systolic array acceleration can effectively increase performance, while maintaining high QoS levels. Up to 44\% system-wide speedups  due to structured pruning and quantization were measured, with only  1.4\% word error rate degradation on the standard LibriSpeech dataset. 

\end{abstract}




\maketitle

\section{Introduction}





Transformers have fostered a revolution in machine learning, with applications ranging from classification~\cite{dosovitskiy2020image} to generative models for text and images~\cite{GPT3}, to speech recognition~\cite{ESPnet}. However, their complex structure based on multiple attention and feed-forward layers~\cite{AttentionIsAllYouNeed} results in unprecedented computational requirements, posing significant challenges for their deployment. 
These are particularly acute in edge scenarios, where systems have to operate within constrained energy and performance envelopes. 

In this context, a plethora of optimization strategies have been proposed. On the software side~\cite{chitty2023survey}, commonly used approaches involve reducing the precision of data representations (quantization) and removing parts that contribute the least to inference outcomes (pruning). As for hardware, efforts have mainly focused on the acceleration of the main computational kernel in transformers, i.e. General Matrix Multiplications (GEMMs).
Although diverse solutions ranging from analog crossbars~\cite{spoon2021toward}\cite{klein2022alpine} to near-DRAM computing~\cite{gomez2021benchmarking}\cite{medina2024bank} have been investigated to this end, a particularly promising alternative is represented by systolic arrays~\cite{SurveySystolicArrayMLAccel}. These two-dimensional meshes of processing elements can indeed parallelize the computation of a GEMM (or, more precisely, the computation of a GEMM tile), while presenting high parallelism, low resource requirements and only mandating a simple, low-overhead control logic. 

Recent works~\cite{Codebench, Scalpel, General_HW_SW_Codesign_Framework_Energy_Efficient_AI, AutomatedHWSWCodesignEdgeAI, Smaug} have attempted to co-optimize software algorithms and hardware accelerators dedicated to transformer inference~\cite{liang2021pruning}. Such a stance is particularly appealing at the crossroads of model pruning and systolic array acceleration. On the software side, pruning can be performed by eliding weights in regular block patterns (in a ``structured" way) rather than as individual elements~\cite{LearningStructuredSparsityInDeepNeuralNetworks}.
While this approach introduces a constraint to pruning, and can hence result in lower overall sparsity rates, it substantially amplifies hardware-side optimization opportunities when matching the sizes of the pruned tile and the accelerator mesh.
The exploration of this strategy, which we term Systolic Array Structured Pruning (SASP), is the focus of this work.

SASP opens a complex multidimensional design space which requires careful consideration of metrics spanning from hardware to algorithms. Indeed, while a larger accelerator can expose a higher degree of parallelism, it also requires more resources (area / energy).
Moreover, SASP settings with larger tiles may overly penalize the achievable sparsity for a desired quality of service (QoS) or, alternatively, result in high QoS degradation for a fixed pruning rate.

To quantitatively explore these interrelations, we present a novel framework and methodology which, for the first time, integrates methods for a) the structured pruning and quantization of transformer algorithms, b) the system-level level modeling of accelerated systems executing them, and c) the hardware synthesis of accelerators. Our environment for SASP exploration builds on state-of-the-art methods for the training of transformers (ESPnet~\cite{ESPnet}) and for system simulation (gem5~\cite{Gem5}). By employing a novel systolic array architectural template, it supports both floating point and weight-quantized data representations, as supported by ESPnet.

As a test case, we employ our exploration approach to analyze speech recognition (ASR) and machine translation (MT) applications,
based on transformers processing the LibriSpeech~\cite{Librispeech} and MuST-C~\cite{digangi2019-mustc} datasets.
We observed that SASP can achieve, for a $32\times32$ systolic array and a ESPnet ASR model trained on LibriSpeech,
up to 44\% speedup and 42\% energy savings over a non-pruned, non-quantized system when employing a 20\% pruning rate, resulting in a marginal Word Error Rate (WER) degradation of 1.4\%.

The contributions of this paper are summarized as follows:

\begin{itemize}
    \item We perform a systematic exploration of Systolic Array Structured Pruning (SASP), a co-design strategy that combines systolic array acceleration and structured pruning with matching accelerator and tile size under quality of service constraints.
    \item We introduce a cross-stack framework to enable the evaluation of SASP through figures of merit at different abstraction levels, including the assessment of QoS, performance, resource usage, and energy, as well as their trade-offs. We discuss how these insights can be effectively leveraged from the joint perspective of algorithmic optimization, system integration, and systolic array design.
    \item Across different models and tasks, we show that SASP-based co-optimization of transformers and systolic arrays can offer significant speedup and energy efficiency gains. For example, a joint ASR and MT case study achieves reductions in runtime and energy consumption of up to 51\% and 34\% while limiting QoS degradation to 4 BLEU points.
\end{itemize}



\section{State of the Art}
\label{sec:background}

By providing spatially-distributed computation with low control logic overhead, systolic arrays can effectively parallelize the execution of matrix multiplications, the dominant computing pattern in transformer inference~\cite{SurveySystolicArrayMLAccel}. To evaluate the benefits that systolic arrays can induce, SMAUG~\cite{Smaug} and TiC-SAT~\cite{TicSat} present system simulation infrastructures able to support complete inferences, including both their hardware-accelerated and their software-executed parts. These works showcase that even small-sized systolic arrays have the potential to reduce run-time by orders of magnitude. 
Performance is further improved when data is properly laid out in a tiled arrangement in memory according to the accelerator characteristics, in order to maximize spatio-temporal locality. Although such an approach has been adopted~\cite{Codebench, amirshahi2024accelerator}, no attempt was made therein to prune computations as we do here with Systolic Array Structured Sparsity.

Indeed, the ductility of DNN models makes them highly amenable to pruning. In the context of systolic array acceleration, pruning optimizations are categorized as either fine-grained or structured~\cite{liang2021pruning}. In the first case, specialized systolic arrays have been proposed which can leverage the presence of zero values in tiles by either clock gating processing elements~\cite{liu2020systolic} or by reordering operands~\cite{tang2024spsa,lu2021sanger}. Nonetheless, fine-grained pruning requires a fair amount of control logic overhead in the accelerator design and impacts the regularity of data layout in memory, which may negate the intended benefits~\cite{ma2021non, Scalpel}.

In this light, structured pruning strategies offer a promising alternative, as tiles of low significance can be entirely skipped before processing them onto accelerators, when the tile size matches the target accelerator parallelism. Hence, speedups can be harnessed without requiring specialized hardware for sparsity management. This position has been adopted by previous works~\cite{sun2023sense, kang2019accelerator, Eridanus, li2020efficient, ben2023variable}. However, they only provide a partial view of the ensuing design space. 
In particular, \cite{sun2023sense, Eridanus} and \cite{kang2019accelerator} adopt a system-level stance, exploring the potential for acceleration of co-designed pruning strategies and data-parallel accelerators, but do not investigate the impact on quality of service (QoS, e.g. accuracy, word error rate) of the performed pruning. 
Conversely, \cite{li2020efficient} and \cite{ben2023variable} provide an algorithmic-level assessment of the effect of structured sparsity, but neglect the hardware and architectural implication of adopting a matched accelerator design. 
To the best of our knowledge, a holistic analysis of the algorithmic-to-hardware space exposed by SASP is hence missing. Our paper aims at filling this gap.

\section{Co-designing Accelerators and Sparsity}
\label{sec:codesign_approach}

\begin{figure}[t]
    \centering
    \includegraphics[width=0.74\linewidth]{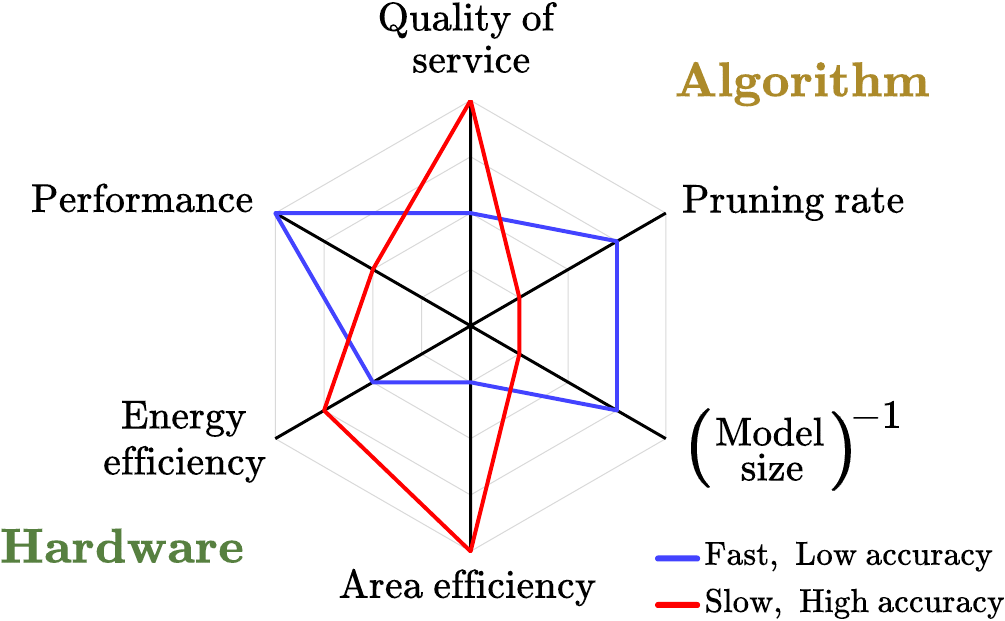}
    \vskip -0.5em
    \caption{Qualitative radar plot illustrating two SASP solutions with different trade-offs: a slow and accurate one (red) employing a small accelerator and a low pruning rate, and a fast but inaccurate one (blue) using a large accelerator and a high pruning rate. Across all axes, higher is better.}
    \label{fig:radar_plot}
\end{figure}

\begin{figure}[tb]
    \centering
    \includegraphics[width=0.97\linewidth,%
    trim={0cm 0.3cm 0cm 0cm},clip%
    ]{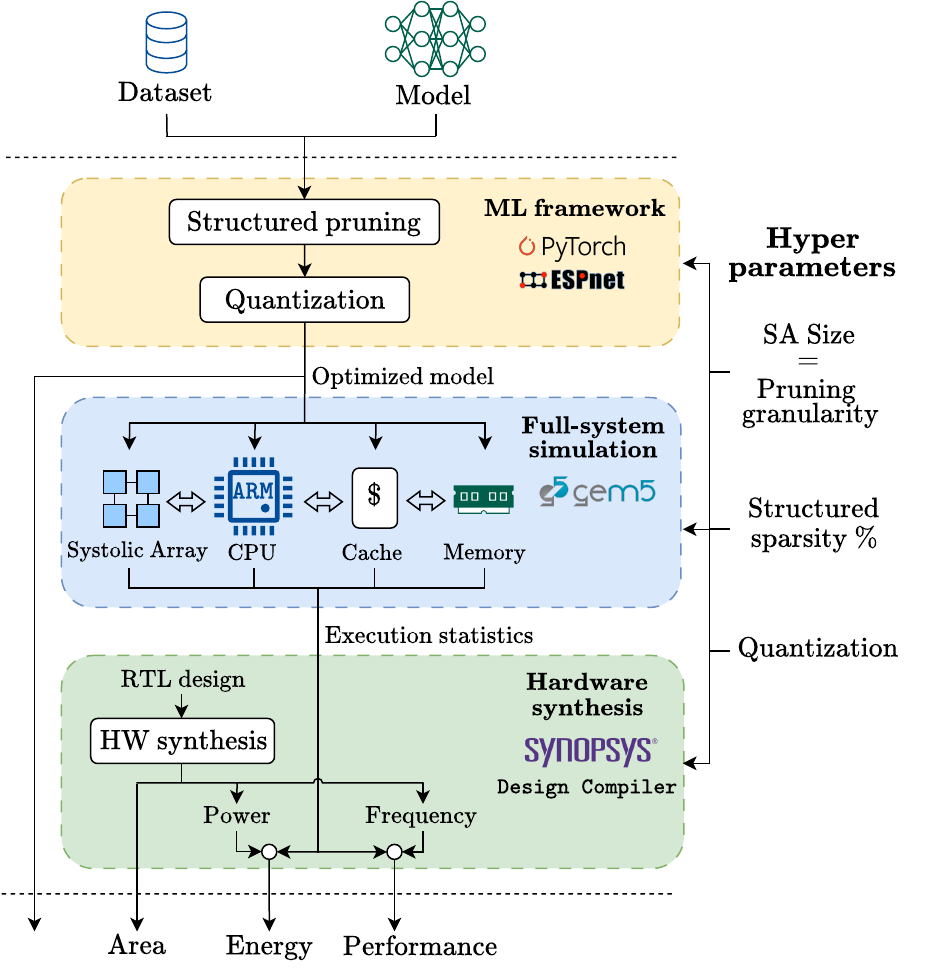}
    \vskip -0.5em
    \caption{Overview of Hardware-Software co-design framework.}
    \label{fig:hw_sw_framework}
\end{figure}


The dimensions of the design space exposed by Systolic Array Structured Pruning (SASP) solutions are illustrated in \Cref{fig:radar_plot}. The illustrated figures of merit reside at widely different layers of the hardware/software stack. Hence, to enable co-optimization, we developed the integrated toolflow illustrated in \Cref{fig:hw_sw_framework}. The input to our framework is a trained transformer model and a target dataset, which, in our implementation, are defined via the ESPnet toolkit for automatic speech processing \cite{ESPnet}. Hyper-parameters determine the size of the SASP tile (which sets the granularity of structured sparsity as well as the size of the systolic array) and the target sparsity rate. Moreover, both floating-point and weight-quantized implementations are supported.

The co-design framework is structured in 3 tiers: in its upper tier, PyTorch APIs are employed to perform pruning and (optionally) quantization. Then, system simulation is employed to gather run-time statistics of the application executing on a virtual system featuring systolic acceleration. Finally, an RTL-level architectural template is used to gather hardware metrics such as energy and area. The implementation of each tier is detailed in the following.


\subsection{Structured Pruning and Quantization}
\label{subsec:pruning_algorithm}

\begin{figure}[tb]
    \centering
    \includegraphics[width=0.8\linewidth,%
    trim={0cm 0.5cm 0cm 0cm},clip%
    ]{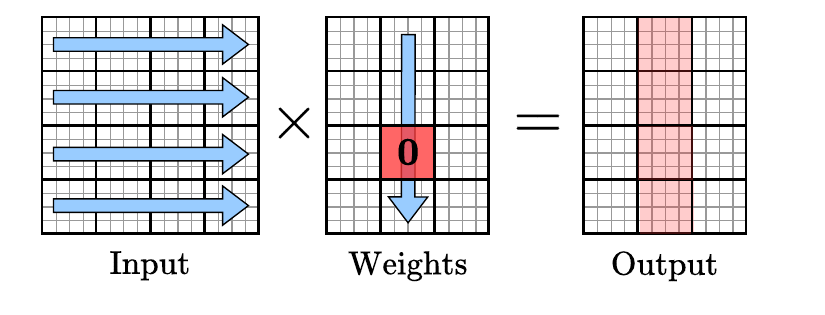}
    \vskip -0.5em
    \caption{Tiled matrix multiplication with structured pruning.}
    \label{fig:tiling_structured_sparsity}
\end{figure}

Matrices employed by transformer models are commonly much larger than the size of systolic arrays. Hence, operations to perform GEMM must be tiled. As in \cite{TicSat}, we herein consider a weight-stationary scenario, in which a tile of parameters is stored in the systolic array, and partial results are computed by streaming inputs/outputs to/from the accelerator. These are eventually aggregated via element-wise addition. 

As shown in \Cref{fig:tiling_structured_sparsity}, in this setting a tile containing only zero values can be completely skipped, saving both the time required to configure the systolic array and the time required to calculate the related partial results. In the example in \Cref{fig:tiling_structured_sparsity} it can be observed that the sparsity induced by the red weight tile lowers the workload required for the computation of the entire shaded column in the output.
We enforce structured sparsity by zeroing a percentage of tiles with the lowest L1-norm (sum of absolute values) across the entire model. This approach allows to heterogeneously  prune GEMMs according to their sensitivity. In particular, feed-forward GEMMs are much more amenable to pruning than attention ones, so we focus on these for our exploration in \Cref{subsec:system_exploration}. 

After sparsification, post-training quantization can optionally be used to reduce the representation precision of weights from 32-bits floating point (FP32) to 8-bits integer (INT8). Finally, inference is  performed on a target dataset, in order to gather QoS metrics such as Word Error Rate (WER) or BLEU. 

\subsection{Full System Simulation}
\label{subsec:fss}

Run-time statistics on the deployment of the SASP-pruned models are collected in the gem5~\cite{Gem5} simulation environment, which allows specifying complex systems including hardware (processors, memory hierarchy) and software (operating system) components. 
To this end, we developed a gem5 custom functional unit modeling a tighly coupled systolic array. Similarly to \cite{TicSat}, the functional unit  employs dedicated instructions, extending the ARM instruction set, to a) program weights, b) perform the systolic array computation, and c) stream inputs/outputs (see \Cref{fig:systolic_array_diagram}). We assume a 32-bit input-output interface, allowing to transfer one input and one output activation per custom instruction. As for weights, either a single FP32 or four INT8 values can be programmed in the array in a quantized or non-quantized setting, respectively.

The implemented instruction set extensions can be employed to accelerate user-level applications via inline assembly pragmas. For convenience, we wrapped these in parametric library functions, allowing to transfer a weight tile or compute a partial GEMM with a single function call. In this way, we gathered the run-time characteristics of executing entire transformer layers under varying architectural and sparsity settings.

\subsection{Systolic Array Architecture}

\label{subsec:systolic_array}

\begin{figure}[t]
    \centering
    \includegraphics[width=0.975\linewidth]{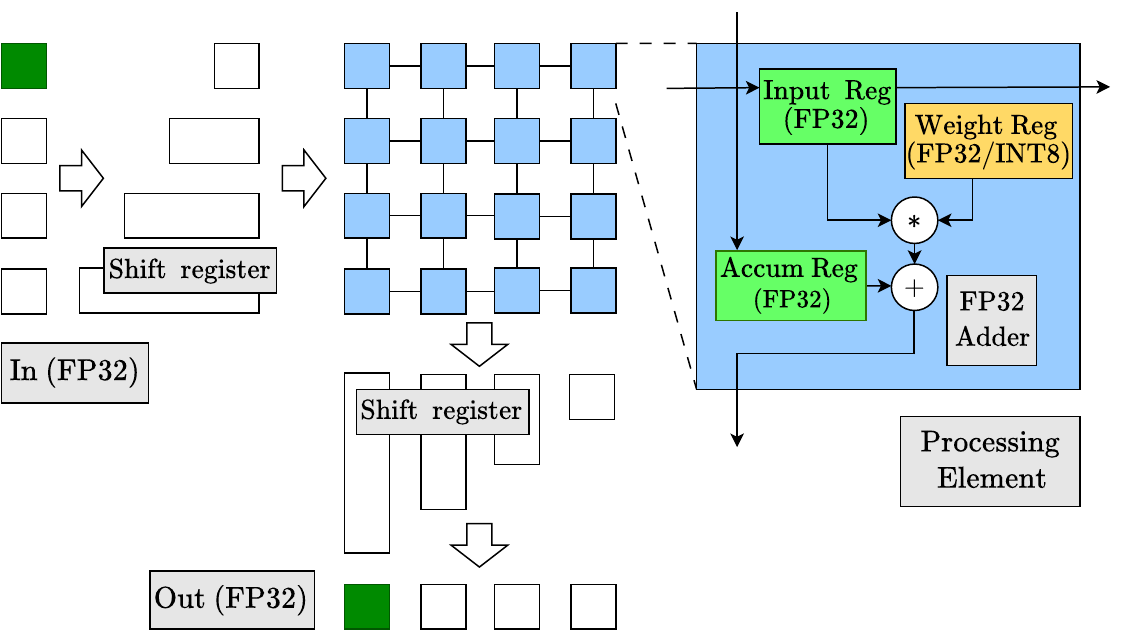}
    \vskip -0.5em
    \caption{Architectural diagram of the systolic array, supporting FP32 activations and either non-quantized (FP32) or quantized (INT8) weights.}
    \label{fig:systolic_array_diagram}
\end{figure}

\begin{figure}[t]
    \centering
    \includegraphics[width=\linewidth]{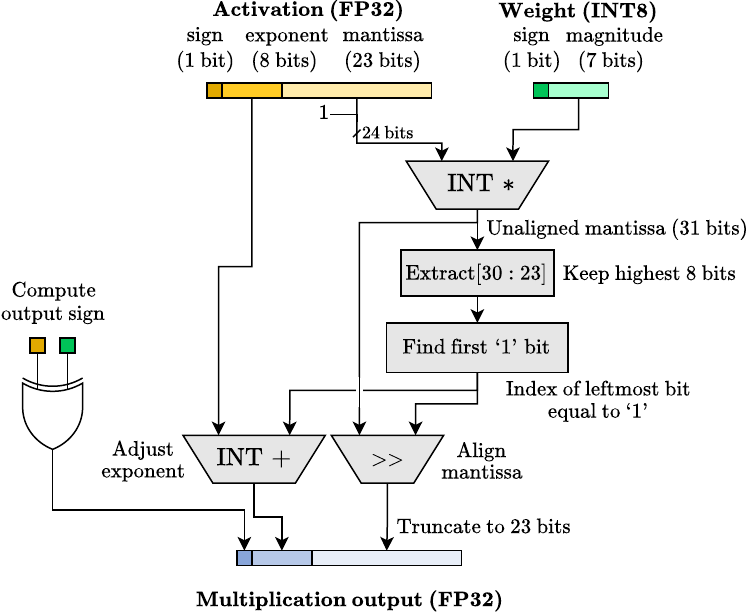}
    \vskip -0.5em
    \caption{Hardware diagram of the hybrid \texttt{FP32\_INT8} multiplier. This logic is bypassed in case any of the operands is equal to zero.}
    \label{fig:fp32_int8_mult}
\end{figure}

The systolic array hardware implementation, depicted in Figure (\Cref{fig:systolic_array_diagram}), comprises a mesh of processing elements (PEs) with nearest-neighbor connections. Inside each, an adder and a multiplier implement a MAC operation between input activation, weight and partial result values, the latter being stored in an accumulation register. Notice that inputs are streamed left-to-right, partial results flow from top to bottom, and weights are instead stationary. At the periphery of the array, shift registers of varying depth are employed to skew data along a diagonal, properly aligning inputs and outputs. 

Instances of the template can be derived by providing architectural parameters defining its size and the desired data format (either FP32 for weights and activations, or using INT8 weights and FP32 activations).
In both versions of the PE, the multiplier and adder are pipelined to meet timing requirements. The pipeline latency is entirely hidden by the activations I/O latency from/to the systolic array. 
The instances are fully synthesizable using standard digital IC design tools and logic cell libraries.

Both in non-quantized and weight-quantized settings, adders in PEs support FP32 representation both for operands and for the results.
Conversely, multipliers can be highly optimized in the weight-quantized case, as they must only support simpler \texttt{FP32\_INT8} arithmetic. 
A diagram  of the hybrid \texttt{FP32\_INT8} multiplier design adopted in our architectural systolic array template is presented in \Cref{fig:fp32_int8_mult}. This implementation correctly computes the multiplication result, except for the case where  either of the inputs equal to 0. We handle this as a special case, by employing a dedicated multiplexer. Moreover, to optimize area and energy efficiency, infinities, NaNs, and subnormal numbers are not handled.

In detail, our design assumes that the INT8 weight is represented using a sign-and-magnitude format. Hence, the output sign is computed as the XOR of the activation and weight signs. Then, the FP32 mantissa is expanded by appending the leading `1', which is implicit in the IEEE format. Furthermore, the expanded mantissa is multiplied by the magnitude of the weight value (INT8).
The resulting unaligned output mantissa is right-shifted to align the leading `1' and truncated to 23 bits. Finally, the output exponent is adjusted according to the number of performed mantissa shifts. 

Note that the hybrid multiplier design readily generalizes to different floating-point and integer bitwidths beyond the \texttt{FP32\_INT8} considered in this paper, e.g., to support FP16 activations.

\section{Experimental Results}
\label{sec:experiments}

\subsection{Setup}
\label{subsed:setup}

\begin{table*}[t]
\caption{Parameters and Quality of Service (QoS) of the deployed models, targeting automatic speech recognition (ASR) and machine translation (MT). Training sets are augmented using 3$\times$ speed perturbation. QoS of the model on the MuST-C dataset~\cite{digangi2019-mustc} are obtained for their execution in cascade (ASR $\rightarrow$ MT).}
\vskip -1em
\label{tab:ESPnetModels}
\begin{center}
\resizebox{\textwidth}{!}{%
\begin{tabular}{@{}ccccccccccccc@{}}
    \toprule
    Toolkit &
    Task &
    Dataset &
    \begin{tabular}{@{}c@{}} Encoder\\blocks \end{tabular} &
    \begin{tabular}{@{}c@{}} Decoder\\blocks \end{tabular} &
    \begin{tabular}{@{}c@{}} Attention\\heads \end{tabular} &
    $d_{model}$ &
    \begin{tabular}{@{}c@{}} Feed-\\forward\\size \end{tabular} &
    \begin{tabular}{@{}c@{}} Training\\set \end{tabular} &
    \begin{tabular}{@{}c@{}} Training\\epochs \end{tabular} &
    \begin{tabular}{@{}c@{}} Test\\set \end{tabular} &
    \begin{tabular}{@{}c@{}} Model\\QoS \end{tabular} &
    \begin{tabular}{@{}c@{}} SASP\\QoS\\target \end{tabular}
    \\\midrule
    ESPnet & ASR & LibriSpeech & 18 & 6 & 4 & 512 & 2048 & 960 h & 100 & 5 h & 3.5\% WER & 5\% WER \\
    ESPnet2 & ASR & LibriSpeech & 12 & 6 & 8 & 512 & 2048 & 960 h & 100 & 5 h & 3.2\% WER & 5\% WER \\
    ESPnet2 & ASR + MT & MuST-C & 18 / 6 & 6 / 6 & 4 / 4 & 128 / 128 & 2048 / 1024 & 560 h & 35 / 200 & 5 h & 31 BLEU & 27 BLEU \\
    \bottomrule
\end{tabular}%
}
\end{center}
\end{table*}

\begin{table}[t]
\setlength{\aboverulesep}{0pt}
\setlength{\belowrulesep}{0pt}
\renewcommand{\arraystretch}{1.15}
\caption{Configuration of the simulated system.}
\vskip -1em
\label{tab:gem5}
\begin{center}
\resizebox{\columnwidth}{!}{%
\begin{tabular}{@{}r|l@{}}
\toprule
Processors & 1x in-order ARMv8 core @1.0 GHz \\
L1-I Cache & 32 kB, 2-way, 2 cycle access \\
L1-D Cache & 32 kB, 2-way, 2 cycle access \\
L2 Cache & 1 MB, 2-way, 20 cycle access \\
Memory & DDR4 2400 MHz, 4 GB \\
Operating System & Ubuntu LTS 16.04 \\
Systolic array & Tightly coupled, control via custom instructions \\
\bottomrule
\end{tabular}%
}
\end{center}
\end{table}

To evaluate our framework's capabilities, we focus on automatic speech recognition and machine translation, two challenging and commercially relevant edge AI tasks, using the LibriSpeech~\cite{Librispeech} and MuST-C~\cite{digangi2019-mustc} corpora. 
Using ESPnet~\cite{ESPnet} and PyTorch~\cite{PyTorch}, we implemented transformer models targeting the encoder for SASP optimization, since its execution dominates run-time. The structure and training method of the models correspond to the parameters shown in \Cref{tab:ESPnetModels}. 
All the quality of service (QoS) results in this paper are reported on the respective corpora test sets and are comparable to state-of-the-art results~\cite{guo_RecentDevelopments_2021}.
System simulations were run in the gem5-X variant \cite{Gem5x} of the gem5 simulator \cite{Gem5}. We considered a single-core  configuration having a 2-level cache hierarchy and running at 1~GHz, as detailed in \Cref{tab:gem5}.
Hardware syntheses of systolic array instances targeted the same 1~GHz timing constraint, and employed a TSMC 28nm technology node. 
Floating point arithmetic operators (adders and multipliers) were derived from the FPxx library using SpinalHDL~\cite{SpinalHDLFP32}, while the hybrid \texttt{FP32\_INT8} multiplier was implemented from scratch according to the design in \Cref{subsec:systolic_array}.

In all experiments, we considered systolic arrays of sizes ranging from
$4\times4$ to $32\times32$. We spanned various structured pruning rates and investigated both \texttt{FP32\_FP32}  and \texttt{FP32\_INT8} quantization schemes. Experiments at different abstraction tiers collected results on area, energy, run-time, and achieved quality of service (QoS). We detail each in the rest of the section. Then, we provide insights from a cross-tier point of view and summarize our findings in \Cref{subsec:cross_tier}. 







\subsection{Hardware Exploration}
\label{subsec:exp-hw}

\begin{figure}[t]%
    \centering
    \subfloat[\centering Area]{{\includegraphics[height=0.475\linewidth]{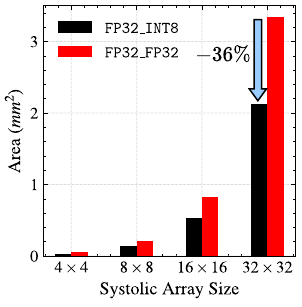} }}%
    \subfloat[\centering Power]{{\includegraphics[height=0.475\linewidth]{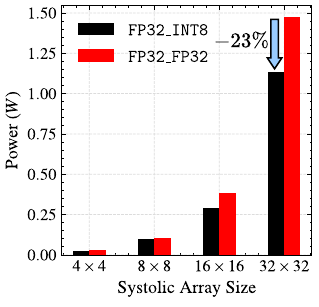} }}%
    \vskip -0.5em
    \caption{Synthesis results for the systolic array design across configurations of varying size (number of rows of the array) and quantization.}
    \label{fig:sa_size_vs_area_power}%
\end{figure}

The area and power results for different systolic array sizes and quantization choices are shown in \Cref{fig:sa_size_vs_area_power}. 
Since multipliers account for an important part of the area and power budget of the entire systolic array (55.6\% and 33.6\%, respectively, in the $8\times8$, \texttt{FP32\_FP32} implementation), the use of the simpler \texttt{FP32\_INT8} design results in tangible savings. In average, these reductions amount to 35.3\% and 19.5\% in area and power across different array sizes. 

Both area and power grow quadratically with the systolic array dimension (e.g. by $\sim$4 times between the $4\times4$ and the $8\times8$ instances), as both the number of PEs and the number of elements in input/output shift registers have a quadratic dependency on the number of array rows/columns.








\subsection{System Exploration}

\label{subsec:system_exploration}

\begin{figure}[t]
    \centering
    \includegraphics[width=\linewidth]{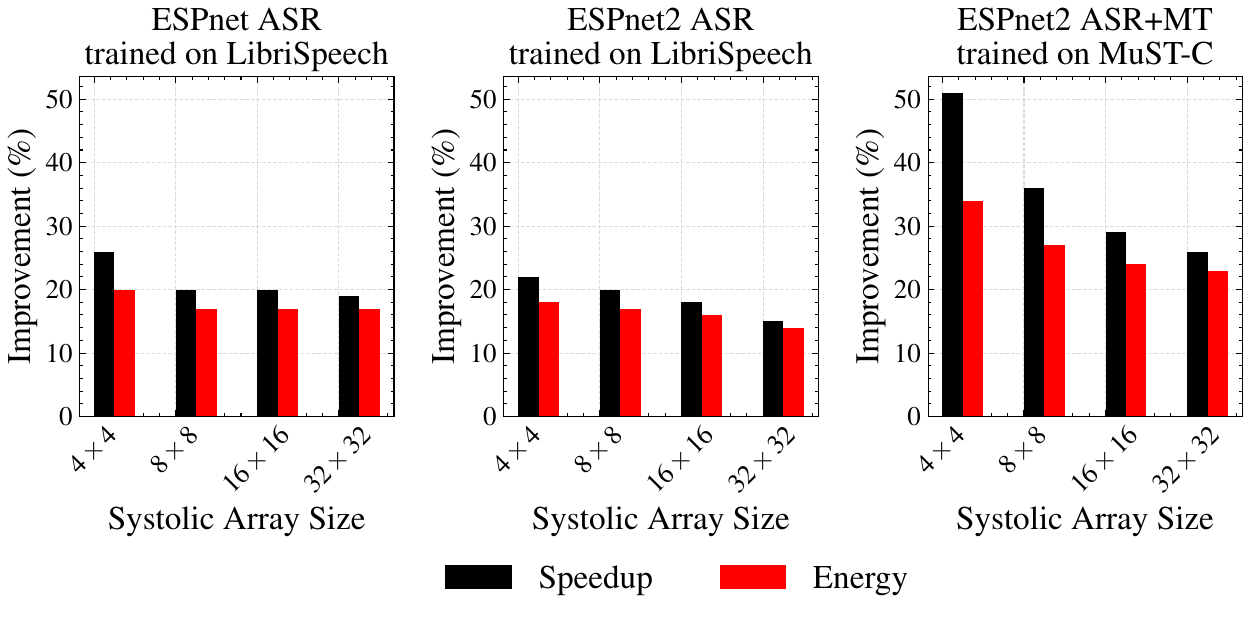}
    \vskip -0.5em
    \caption{
    Speedup and energy improvements from Systolic Array Structured Pruning under the quality of service target in \Cref{tab:ESPnetModels} with respect to non-pruned quantized executions, considering  systems with systolic array accelerators of varying size. Results are shown for \texttt{FP32\_INT8} systolic array configurations.
    }
    \label{fig:speedup_energy_improvements}
\end{figure}





We first assess the overall impact of SASP across workloads and systolic array sizes on runtime and energy consumption under the target QoS degradations defined in \Cref{tab:ESPnetModels}. As shown in \Cref{fig:speedup_energy_improvements}, speedup and energy improvement opportunities depend on the workload due to their different sensitivities to pruning. For example, maximum speedup / energy improvements vary from 26\% / 21\% in ESPnet ASR on LibriSpeech, to 22\% / 18\% in ESPnet2 ASR on LibriSpeech, to 51\% / 34\% in ESPnet2 ASR+MT on MuST-C. However, their trends when changing accelerator size are consistent across workloads, e.g., achievable improvements decrease when employing larger systolic arrays. To thoroughly analyze the trade-offs among encoder inference speedup, area-energy product and quality of service, all further explorations will focus on the ESPnet automatic speech recognition model using the LibriSpeech corpus.

\begin{figure}[t]
    \centering
    \includegraphics[width=\linewidth]{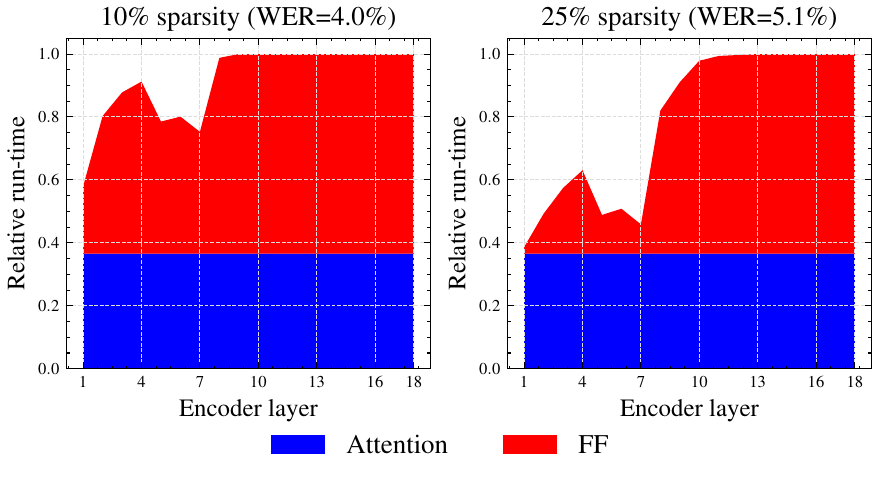}
    \vskip -0.5em
    \caption{
    Per-layer normalized run-time of the transformer encoder after applying Systolic Array Structured Pruning at two global sparsity targets. Results are shown for an $8\times8$, \texttt{FP32\_INT8} systolic array, with the execution time of each layer normalized to the execution without pruning.
    }
    \label{fig:per_layer_sparsity_and_run-time}
\end{figure}

\Cref{fig:per_layer_sparsity_and_run-time}~plots the measured per-layer normalized encoder run-time of a systolic-accelerated system performing an inference. Data refers to an $8\times8$ systolic array, with varying degrees of structured sparsity. Speedup numbers closely follow sparsity levels, as inference run-time is  strongly dominated by GEMM computations (exceeding 97\% in all cases~\cite{TicSat}). 

Pruning was performed according to the methodology proposed in Section \ref{subsec:pruning_algorithm} targeting feed-forward layers, which exhibit a much higher degree of resilience and account for the largest part of the workload. Results in \Cref{fig:per_layer_sparsity_and_run-time} highlight that early feed-forward layers are the most amenable to pruning, while later ones have a higher proportion of tiles with a non-negligible L1-norm, which have a higher impact on inference outcomes.


\subsection{Impact of SASP on Quality of Service}
\label{subsec:sasp_vs_qos}

\begin{figure}[t]%
    \centering
    \subfloat[\centering \texttt{FP32\_FP32}]{{\includegraphics[width=0.485\linewidth]{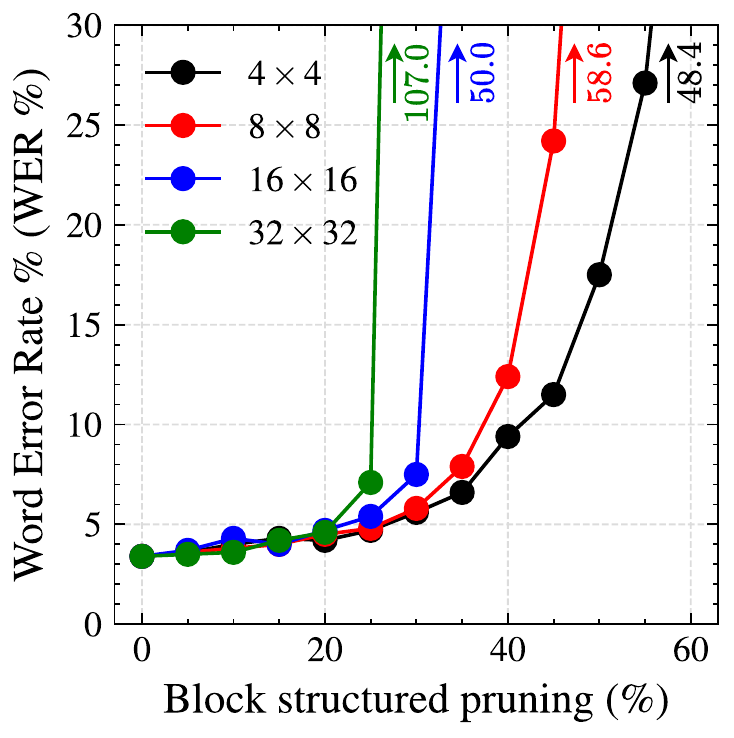} }}%
    \subfloat[\centering \texttt{FP32\_INT8}]{{\includegraphics[width=0.485\linewidth]{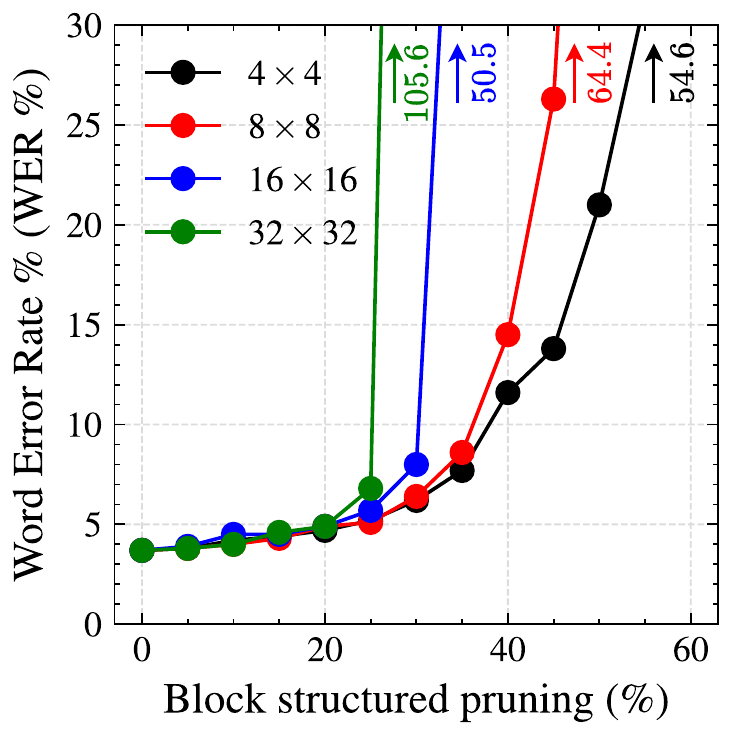} }}%
    \vskip -0.5em
    \caption{Achieved Word Error Rate when varying the percentage of Systolic Array Structured Pruning.}
    \label{fig:WER_vs_pruning_percentage_per_sa_size_fp32_int8}%
\end{figure}

\Cref{fig:WER_vs_pruning_percentage_per_sa_size_fp32_int8} shows that Word Error Rate (WER) grows exponentially when increasing the degree of Systolic Array Structured Pruning. Similar trends are present for both the weight-quantized model (indicated as \texttt{FP32\_INT8}) and the non-quantized one (\texttt{FP32\_FP32}).

As the size of the systolic array grows, trends become steeper, showcasing an abrupt increase in WERs at smaller SASP rates. This effect is caused by the higher brittleness of large-tile structured pruning  with respect to small-tile cases. Indeed, while it may be possible to find four prunable $4\times4$ tiles (containing 64 values in total), selecting a single contiguous $8\times8$ tile (again, of 64 values) can be considerably more challenging.

\subsection{Multidimensional SASP Trade-offs} 

\begin{figure}[t]
    \centering
    \includegraphics[width=\linewidth]{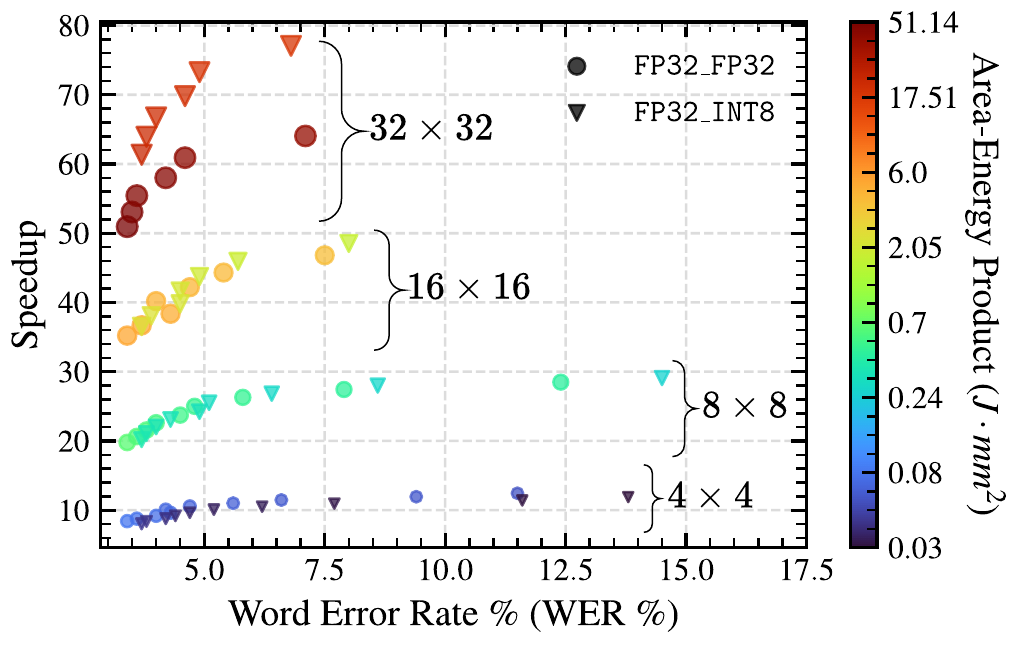}
    \vskip -0.5em
    \caption{Trade-offs among encoder inference speedup, area-energy product and Word Error Rate across systolic array sizes and structured pruning rates. Speedup is computed with respect to a non-accelerated, non-quantized baseline.}
    \label{fig:bubblechart_WER_perf_AEP}
\end{figure}

Focusing again on speech recognition, \Cref{fig:bubblechart_WER_perf_AEP} shows the variations in performance, QoS and resource usage when changing SASP rate, quantization strategy, and systolic array size. WER and Speedup are plotted on the two axes, the marker shape discriminates between FP32 and weight-quantized implementations, and the marker colors indicate their resource requirements in terms of Area-Energy product.  Data points form four distinct clusters corresponding to each systolic array size. Two curves in each cluster represent the two quantization choices, which have notably different Area-Energy products. The non-quantized \texttt{FP32\_FP32} version achieves lower (better) WER, with the differences becoming more pronounced at higher pruning rates and for larger systolic arrays.

From a run-time perspective, \texttt{FP32\_INT8} configurations allow to reduce the cost of weight transfers by loading four INT8 weights per 32-bit bus access, as opposed to a single FP32 one in the \texttt{FP32\_FP32} configuration. Consequently, \texttt{FP32\_INT8} implementations outperform their \texttt{FP32\_FP32} counterparts for systolic array sizes larger than $4\times4$, as the savings in data transfers offset software/system overhead.
Nonetheless, while quantization has a large impact on area and energy, its influence on performance is smaller, as the majority of the run-time is not spent in weight data transfers, but instead for streaming inputs / computing outputs, which is equally fast for both quantization schemes.

Within each systolic array size and quantization configuration, the SASP pruning rate guides the trade-off between inference time and QoS. \Cref{fig:bubblechart_WER_perf_AEP} shows that, up to an inflection point at a WER of $\sim$5\%, SASP enables to strike effective fine-grained balances between run-time performance and QoS. Beyond this inflection point, further increases in  pruning rates cause instead very high WER degradations. 

\Cref{tab:tradeoffs_wer_5} illustrates the effect of applying Systolic Array Structured Pruning and weight quantization at the 5\% inflection point. In this setting, SASP improves performance and energy consumption up to 26\% and 21\%, respectively. Furthermore, when combining quantization and structured pruning, performance and energy efficiency improvements  reach 44\% and 42\%, while also decreasing area occupation by 36\%.
These substantial gains are achieved without increasing the systolic array size, and hence do not incur the hefty associated 
area and energy costs, as also depicted in \Cref{tab:tradeoffs_wer_5}. 
As an example, scaling from an $8\times8$ to a $32\times32$ systolic array does yield a 3.04$\times$ speedup for \texttt{FP32\_INT8} quantization, but also requires 15.21$\times$ more area and 3.98$\times$ more energy.

\begin{table}[t]
\setlength{\aboverulesep}{0pt}
\setlength{\belowrulesep}{0pt}
\renewcommand{\arraystretch}{1.15}
\caption{Area, encoder speedup and energy results for different systolic array configurations without SASP (3.5$\pm$0.2\% WER) and with SASP (5$\pm$0.4\% WER). Speedup is computed with respect to a non-quantized baseline executed on CPU.}
\vskip -1em
\label{tab:tradeoffs_wer_5}
\begin{center}
\resizebox{\columnwidth}{!}{%
\begin{tabular}{@{}r|c|c|cccc@{}}
\toprule
   \multicolumn{3}{c|}{Size} & $4\times4$ & $8\times8$ & $16\times16$ & $32\times32$\\\bottomrule
   \multirow{6}{*}{\texttt{FP32\_FP32}} & \multicolumn{2}{c|}{Area $(mm^2)$} & 0.05 & 0.21 & 0.83 & 3.34 \\\cline{2-7}
   & No & Speedup & 8.42 & 19.79 & 35.22 & 50.95 \\\cline{3-7}
   & SASP & Energy $(J)$ & 1.60 & 3.09 & 6.37 & 15.32 \\\cline{2-7}
   & & Pruning (\%) & 25 & 25 & 20 & 20 \\\cline{3-7}
   & \textbf{SASP} & Speedup & 10.56 & 25.01 & 42.21 & 60.91 \\\cline{3-7}
   & & Energy $(J)$ & 1.27 & 2.43 & 5.28 & 12.70 \\\cline{3-7}\bottomrule
  \multirow{6}{*}{\texttt{FP32\_INT8}} & \multicolumn{2}{c|}{Area $(mm^2)$} & 0.03 & 0.14 & 0.53 & 2.13 \\\cline{2-7}
   & No  & Speedup & 8.03 & 20.18 & 36.53 & 61.33 \\\cline{3-7}
   & SASP & Energy $(J)$ & 1.24 & 2.67 & 4.57 & 10.64 \\\cline{2-7}
   & & Pruning (\%) & 25 & 20 & 20 & 20 \\\cline{3-7}
   & \textbf{SASP} & Speedup & 10.08 & 24.23 & 43.74 & 73.25 \\\cline{3-7}
   & & Energy $(J)$ & 0.99 & 2.21 & 3.79 & 8.82 \\\bottomrule
   
\end{tabular}%
}
\end{center}
\end{table}


\subsection{Cross-tier Analysis}
\label{subsec:cross_tier}

\begin{figure}[t]%
    \centering
    \subfloat[\centering \texttt{FP32\_FP32}]{{\includegraphics[width=0.485\linewidth]{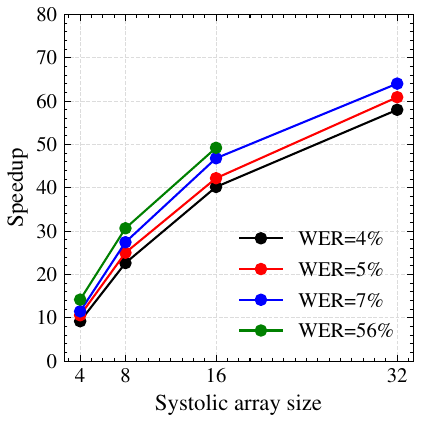} }}%
    \subfloat[\centering \texttt{FP32\_INT8}]{{\includegraphics[width=0.485\linewidth]{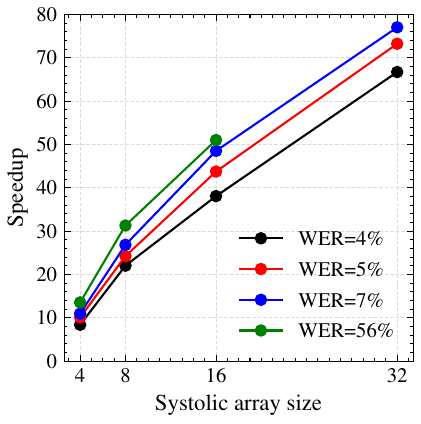} }}%
    \vskip -0.5em
    \caption{Speedup with respect to software execution of the encoder while varying the systolic array size, for different Word Error Rates. 
    }
    \label{fig:sa_size_vs_speedup_constant_wer}%
\end{figure}

As discussed above, increasing the systolic array size increases run-time performance by offering higher parallelism, but also restricts the achievable pruning rate for a given QoS. Therefore, when using SASP, increases in the array size (hence the tile size in GEMM computations) result in diminishing gains for a given WER target, as the pruning rate must be lowered to maintain QoS. 
This trend is illustrated in \Cref{fig:sa_size_vs_speedup_constant_wer}, which highlights a sublinear relation between systolic array size and speedup, both in the case of FP32 and weight-quantized scenarios. Hardware costs (area / energy), instead, increase quadratically with size, as outlined in \Cref{subsec:exp-hw}, requiring careful co-design considerations, especially when targeting resource-constrained edge systems.

\section{Conclusion}
Deploying transformers on edge devices requires both software optimization and hardware acceleration to meet strict resource constraints while maintaining performance.
In this paper, we have analyzed their interaction, focusing on structured sparsity and systolic arrays. We explored Systolic Array Structured Pruning (SASP), where  the size of pruned blocks is matched to the dimensions of the systolic array, enabling the skipping of entire computation tiles. 
To assess the benefits and pitfalls of SASP, we presented a cross-stack framework to co-optimize edge AI transformers, which integrates algorithmic optimization, system simulation, and hardware design. Employing it, we performed a comprehensive analysis of how SASP, quantization, and systolic array configurations affect area, energy, performance, and quality of service (QoS) in edge transformers for speech recognition and machine translation.

Our results demonstrate that SASP provides fine-grained control over the trade-off between inference run-time and QoS up to an inflection point, after which increased pruning drastically degrades QoS for small performance gains. 
Experimental evidence on the ESPnet ASR transformer trained on LibriSpeech has shown that system-wide performance improvements of up to 44\% and accelerator energy reductions of up to 42\% can be obtained under a 1.4\% WER degradation, when employing weight quantization and a 20\% pruning rate.
Additionally, we showcased that, although larger systolic arrays do reduce run-time, they also incur substantial energy and area costs, while yielding sublinear speedups for a target QoS. This sublinearity emerges from the reduced structured pruning opportunities, as finding contiguous zero blocks becomes harder with increasing block sizes. For this reason, SASP is particularly well-suited for edge AI accelerators, where stringent resource constraints are present.

\begin{acks}
This work was supported in part by the Swiss State Secretariat for Education, Research, and Innovation (SERI) through the SwissChips Research Project; in part by the Swiss NSF Edge-Companions project (GA No. 10002812); in part by the EC H2020 FVLLMONTI Project under Grant 101016776; and in part by ``La Caixa'' Foundation (ID 100010434) under Grant LCF/BQ/EU23/12010073. This research was partially conducted by ACCESS – AI Chip Center for Emerging Smart Systems, supported by the InnoHK initiative of the Innovation and Technology Commission of the Hong Kong Special Administrative Region Government.
\end{acks}

\bibliographystyle{ACM-Reference-Format}
\bibliography{references}










\end{document}